\documentclass[reprint,amsmath,amssymb,aps,prl,groupedaddress,showpacs]{revtex4-1}

\usepackage{graphicx}% Include figure files
\usepackage{dcolumn}% Align table columns on decimal point
\usepackage{bm}% bold math
\usepackage{color}
\begin{document}

\title{Self-consistent calculations of the electric giant dipole resonances\\ in light and heavy mass nuclei.}

\author{N. Lyutorovich}
\author{V. Tselyaev}
\affiliation{V. A. Fock Institute of Physics, St. Petersburg State University,
RU-198504 St. Petersburg, Russia}
\author{J. Speth}
\email{J.Speth@fz-juelich.de}
\author{S. Krewald}
\author{F. Gr\"ummer}
\affiliation{Institut f\"ur Kernphysik, Forschungszentrum J\"ulich, D-52425 J\"ulich, Germany}
%\author{S. Kamerdzhiev}
%\affiliation{Institute of Physics and Power Engineering, 249020 Obninsk, Russia}
\author{P.-G. Reinhard}
\affiliation{Institut f\"ur Theoretische Physik II, Universit\"at Erlangen-N\"urnberg,
D-91058 Erlangen, Germany}

\date{\today}

\begin{abstract}
While bulk properties of  stable nuclei are successfully reproduced 
by mean-field theories employing effective interactions, 
the dependence of the centroid energy of the electric giant dipole resonance 
on the nucleon number A is not.  This problem  is cured by considering
many-particle correlations beyond mean-field theory, which we do 
within the \emph{Quasiparticle Time Blocking Approximation}.
The electric giant dipole resonances in $^{16}$O, $^{40}$Ca, 
and $^{208}$Pb are calculated using two new Skyrme interactions.
\end{abstract}


\maketitle

%\section{Introduction}

The electric giant dipole resonance (GDR) is a well-known nuclear excitation 
mode which is related to bulk properties of nuclei,
such as  the Thomas-Reiche-Kuhn (TRK) sum rule and 
the nuclear symmetry energy \cite{Berman:1975tt}.
One might assume that theories which describe both bulk properties of nuclei 
and shell effects rather well, such as self-consistent mean-field theories
based on effective nucleon interactions 
\cite{Niksic:2011sg,Bender:2003jk,Goriely:2002xz,Kortelainen:2010hv},
should have no problem in systematically reproducing the centroid energies 
of the GDR as a function of the nucleon number A. 
This is not the case, however, as has been discussed in detail in several 
recent reviews on mean-field theories which include strength functions 
obtained within the quasi-particle random-phase approximation (QRPA) 
\cite{Klupfel:2008af,Erler:2011,Erler:2010fe,Dutra:2012mb}. 
It was impossible so far to describe ground-state properties and the
centroid energy of the GDR both in light and heavy nuclei with the same
effective interaction.
The problem is more serious than might appear at a first glance because 
the physics of the GDR is intimately related to the neutron skin thickness 
and the pygmy dipole strength 
\cite{Abrahamyan:2012,Tamii:2011pv,Savran:2011zza},  
presently investigated experimentally because of an impact on the isotope 
abundance produced in supernova explosions \cite{Horowitz:2000xj}.
There are two hints suggesting that the mean field approach 
by itself is at the origin of the problem. Complex configurations play 
a well-known role in the damping of nuclear excitations \cite{Bertsch:1983zz}.
Even when effective interactions are fitted to the effective isoscalar mass,
the symmetry energy, and the TRK sum rule enhancement factor $\kappa$, 
the problem remains unsolved \cite{Dutra:2012mb}.

We employ the \emph{Quasiparticle Time Blocking Approximation} (QTBA),  
developed and applied in 
\cite{Tselyaev:1989,Tselyaev:2005de,Kamerdzhiev199390,Avdeenkov2007196,Lyutorovich:2008nv,Tselyaev_Speth:2007,Kamerdzhiev:2003rd}, 
to study the GDR. The QTBA is a method to calculate nuclear response functions 
which generalizes the QRPA.  It includes explicitly
the coupling of one-particle one-hole(1p1h) configuration with phonons,  
but omits the simultaneous excitation of two-phonon states in the presence 
of a 1p1h-excitation. In the limit of vanishing phonon-nucleon coupling, 
the QTBA corresponds to the QRPA, a standard mean field approach. 
Originally, the QTBA  was used in the framework of Landau-Migdal theory, 
but has been generalized recently to effective interactions of the Skyrme family 
in order to make possible self-consistent calculations \cite{Avdeenkov2007196,Lyutorovich:2008nv,Tselyaev2009}. 
The Skyrme interactions are defined by a set of momentum- and density-dependent 
contact interactions; different parameterizations may be distinguished 
by some set of theoretical quantities, such as nuclear matter properties 
or the effective mass, which are not directly observable. 
The momentum dependence of the Skyrme interaction leads to an effective mass, 
with values $m^*/m < 1$ found by many investigations. 
Mean-field approaches which employ effective masses smaller than unity 
generate single-particle energies which systematically deviate 
from the separation energies, mainly by a too small level density.
Larger level densities can be obtained by taking into account 
the energy dependence of the nucleon self-energy, as is shown 
in Refs.~\cite{Brown1963598,Bertsch1968204,Ring1973198,Jeukenne:1976uy}. 
The energy dependence of the self-energy is due to complex configurations,
such as the coupling of phonons to the single particle degrees of freedom. 
 In this communication we show that if these effects are considered,
 both centroid energies  and  spreading widths of the giant resonances are reproduced.
As we know that Skyrme forces cannot reproduce simultaneously 
the GDR in $^{16}$O and $^{208}$Pb \cite{Erler:2010fe}, 
we have adjusted new Skyrme parameterizations for the purpose 
of this study concentrating on tuning the GDR in $^{16}$O  within 
the mean-field approach (RPA). Since there are only few collective 
nuclear vibrations in light nuclei,
the inclusion of phonons within the QTBA is expected to produce results
close the ones obtained in the  mean field approach for $^{16}$O.
On the other hand, in heavy nuclei, the number of collective modes increases,
which leads to major differences between the mean-field approach and the QTBA.
We follow exactly the same fitting strategy and data as used for the systematic 
variation of forces in \cite{Klupfel:2008af}.
As a result we obtain two new forces, SV-m56-O with effective 
mass $m^*/m=0.56$ and SV-m64-O with $m^*/m=0.64$.
Both forces have a rather low symmetry energy $a_\mathrm{sym}=27$ MeV, 
and high sum-rule enhancement factor \cite{Klupfel:2008af} 
$\kappa_\mathrm{TRK}=0.6$. The parameters are listed in Table~\ref{tab:force}.

\begin{table}[h!]
\caption{\label{tab:force}
Skyrme force parameters (upper block), adjusted nuclear matter
properties (middle block), and dipole polarizability
$\alpha_\mathrm{D}$
as well as neutron skin $r_\mathrm{rms,n}-r_\mathrm{rms,p}$ (lower
block) for the two newly designed Skyrme forces.
The standard force parameters are given where $\alpha$ is the power of the
density dependence.
All three
forces use Coulomb exchange in Slater approximation and the
c.m. energy correction $\langle\hat{P}_\mathrm{cm}^2\rangle/(2mA)$.
For details of the functional and options see
\cite{Bender:2003jk,Klupfel:2008af,Erler:2011}.
}
\begin{ruledtabular}
\begin{tabular}{c|rr}
  & SV-m56-O & SV-m64-O  \\
\hline
 $t_0$ & -1905.403 & -2083.855 \\
 $t_1$ &  571.187 &   484.604 \\
 $t_2$ & 1594.803 & 1134.345  \\
 $t_3$ & 8439.036 & 10720.663  \\
 $t_4$ & 133.268 &  113.973  \\
 $x_0$ & 0.644020 & $\; \;$ 0.619768  \\
 $x_1$ & -2.973738 & -2.332678   \\
 $x_2$ & -1.255261 & -1.305938   \\
 $x_3$ & 1.796625 & 1.210109 \\
 $b'_4$ &  52.97011 & 62.92567 \\
 $\alpha$ & 0.2    & 0.2     \\
 $\frac{\hbar^2}{2m_p}$ & 20.74982   & 20.74982     \\
 $\frac{\hbar^2}{2m_n}$ &  20.72126  &  20.72126   \\
\hline
 $m^*/m$  & 0.56  & 0.64   \\
 $a_\mathrm{sym}$/MeV & 27 & 27 \\
 $\kappa_\mathrm{TRK}$ & 0.6 & 0.6 \\
\hline
$\alpha_\mathrm{D}$/fm$^3$ &  20.2 & 19.4  \\
n-skin [fm] &   0.156 & 0.134  \\
\end{tabular}
\end{ruledtabular}
\end{table}

 Of course, the new fits maintain the good ground state properties
of all the systematically varied forces in [6]. Additionally, the low effective mass and low symmetry energy $a_\mathrm{sym}$=27 MeV 
together with a rather high sum-rule enhancement factor
$\kappa_\mathrm{TRK}$ of the two parameterizations delivers a high GDR energy. 
This is beneficial for $^{16}$O but leads to its overestimate in $^{208}$Pb. 
However the inclusion of complex configurations brings the GDR in $^{208}$Pb 
down to the correct value.
Similar effects are seen for the single-particle energies. The parameterizations 
SV-m56-O and SV-m64-O reproduce these energies in $^{16}$O 
reasonably well, but the single-particle spectrum in $^{208}$Pb 
is spread out too much and deviates strongly from the experimental one. 
The coupling to the phonons will improve the spectrum \cite{Tselyaev_Speth:2007}.
Being built on RPA, QTBA follows basically the same trends with
varying Skyrme force as RPA (see [6]),though the absolute value of the effect is different in light and heavy mass nuclei. 

 Let us outline some technical details of our numerical scheme.
In our RPA and QTBA calculations of the GDR, the single-particle continuum
is treated exactly according to the scheme described in Ref. \cite{Lyutorovich:2008nv}.
The phonons were calculated within the so-called discretized RPA (DRPA).
Here the results depend on the single particle basis and on details 
of the discretization e.g. the size of the box one chooses. 
In the present investigation, such ambiguities are small, as we control 
them by comparing the DRPA results with a full continuum RPA.

 In self-consistent calculations the {\it ph}-interaction is given 
by the second derivative of the energy functional.
In the DRPA calculations of phonons the matrix elements of the ph-interaction 
were calculated exactly except for the spin-orbit and 
Coulomb contributions which were omitted.
In the RPA and QTBA calculations of the GDR we used additional 
(local-exchange) approximation for the velocity-dependent part 
of the ph-interaction derived from Skyrme energy functional.
In the case of the GDR this approximation gives results 
which are close to the exact RPA results.
It will be described in a forthcoming publication.
In our investigation we did not study spin-dependent properties of new Skyrme forces.
It is well-known that spin-spin part of the residual interaction
(except for ${\mbox{\boldmath $J$}}^2$-generating terms) does not contribute
to the ground-state structure for spherical even-even nuclei. 
So one can omit this part in the calculations
of the excited states in these nuclei without breaking self-consistency.
On the other hand, inclusion of the spin-spin part of the residual {\it ph}-interaction
leads to an instability in the DRPA calculations of the phonon's characteristics with
given Skyrme forces. In fact, as for most other Skyrme forces, the instability is driven by
the term $\propto(\nabla\sigma)^2$ in the functional. For this reason we exclude this part of the interaction in our
calculations.

The number of solutions of the RPA equations depends on the size 
of the configuration space. However, the majority of the RPA wave functions 
are dominated by one particle-hole configuration.
In principle, one has  to subtract the second order contributions to complex (ph-phonon) configurations in order to avoid 
double counting \cite{Leieune:1979}.
For simplicity, the present calculations consider only a small 
number of  phonons,  defined by having transition probabilities of 
at least 1/5 of the strongest state of each multipolarity.
For these phonons, the second-order corrections are small and 
have been neglected.
\begin{figure}[htbp]
\vskip 4mm
\begin{center}
\includegraphics[width=8cm]{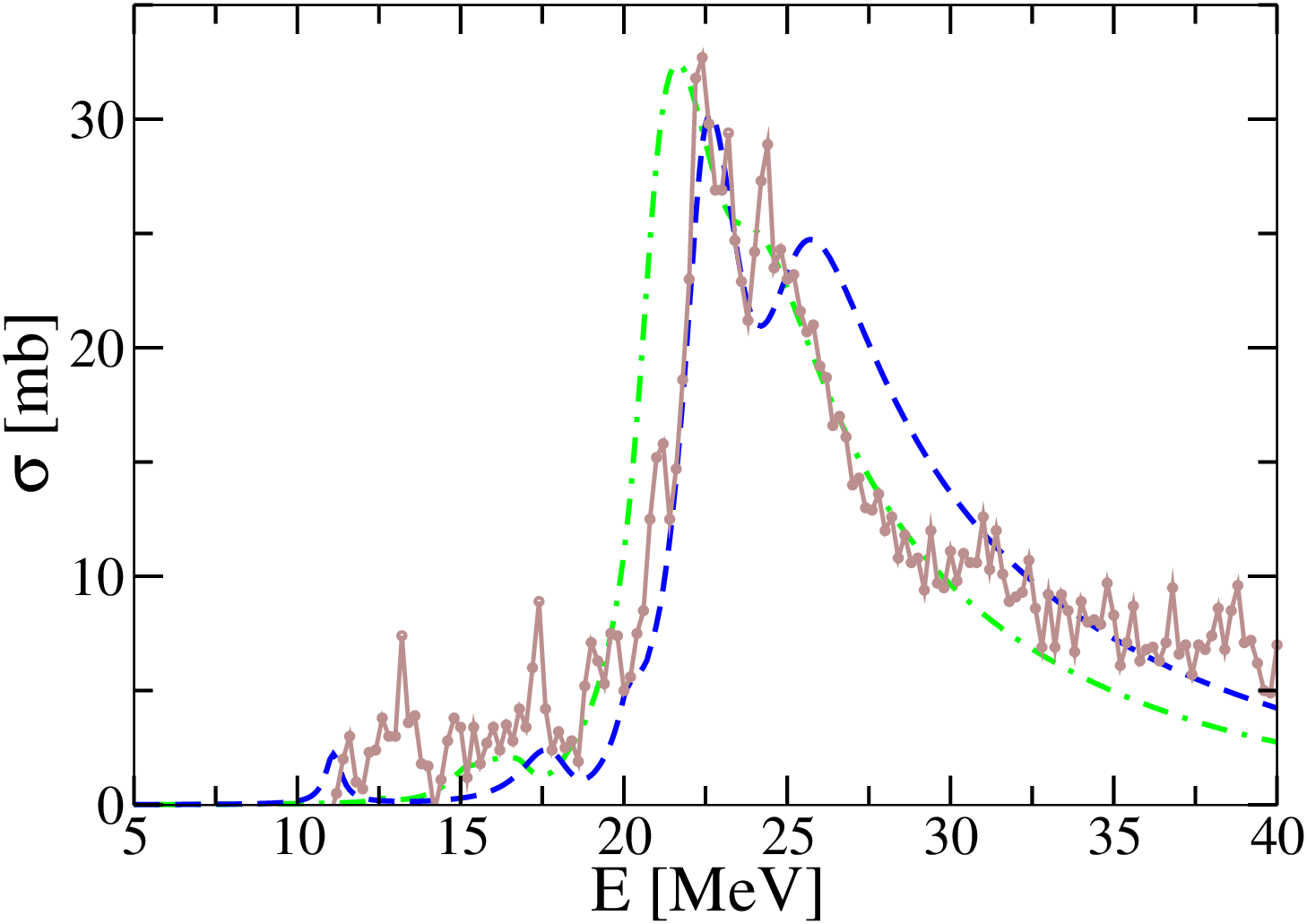}
\end{center}
\caption{\label{fig:1} (Color online) Photo absorption cross-section in $^{16}$O 
calculated self-consistently in RPA, using two different Skyrme parameterizations 
with effective mass 0.56 (dashed(blue)) line) and 0.64 (dashed-dotted(green) line). 
The experimental cross section are given by the (brown) dots connected 
by a solid line \cite{Ishkhanov:2002}. }
\end{figure}
\begin{table}[h]
\caption{Comparison of  theoretical and experimental
\cite{Belyaev,Erokhova,Ishkhanov:2002} Lorentzian parameters.
The  energies considered range from 8-25 [MeV] for $^{208}$Pb 
and from 10-32 [MeV] for $^{40}$Ca and $^{16}$O. }
\label{tab:Lorentz}
\begin{ruledtabular}
\begin{tabular}{llccr}
Nucleus & Force & $\overline{E}$ [MeV] & $\Gamma$ [MeV] & $\sigma_0$ [mb] \\
\hline
$^{208}$Pb & SV-m56-O (RPA)  & 14.30 & 4.96 & 624 \\
           & SV-m56-O (QTBA) & 13.37 & 5.99 & 495 \\
& Experiment                 & 13.43 & 5.08 & 481 \\
\hline
$^{40}$Ca  & SV-m56-O (RPA)  & 21.61 & 5.90 & 104 \\
           & SV-m56-O (QTBA) & 21.14 & 5.92 &  99 \\
& Experiment                 & 20.00 & 5.00 &  95 \\
\hline
$^{16}$O   & SV-m56-O (RPA)  & 25.31 & 8.95 & 25.7 \\
           & SV-m56-O (QTBA) & 24.49 & 8.85 & 24.8 \\
& Experiment                 & 23.76 & 7.17 & 24.8 \\
\end{tabular}
\end{ruledtabular}
\end{table}
%\section{Results}

%
\begin{figure}[htbp]
\vskip 4mm
\begin{center}
\includegraphics[width=8cm]{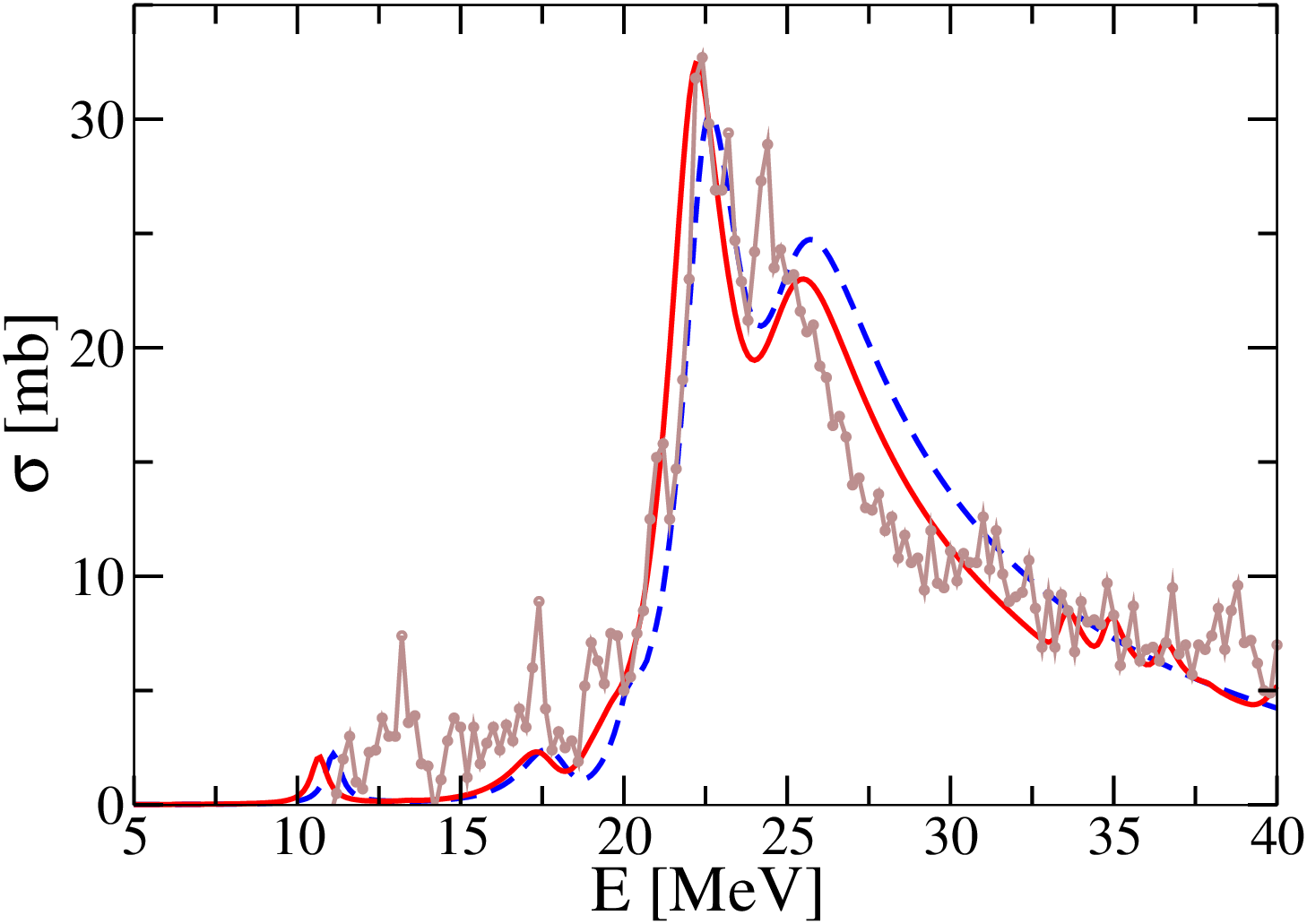}
\end{center}
\caption{\label{fig:2} (Color online) Comparison of the experimental 
\cite{Ishkhanov:2002} photo absorption cross-section in $^{16}$O 
with theoretical ones calculated in RPA (dashed(blue) line) and 
QTBA (solid(red) line) using the SV-m56-O Skyrme parameters.
The experimental data are given by the (brown) dots connected 
by a solid line \cite{Ishkhanov:2002}.  }
\end{figure}
\begin{figure}[htbp]
\vskip 4mm
\begin{center}
\includegraphics[width=8cm]{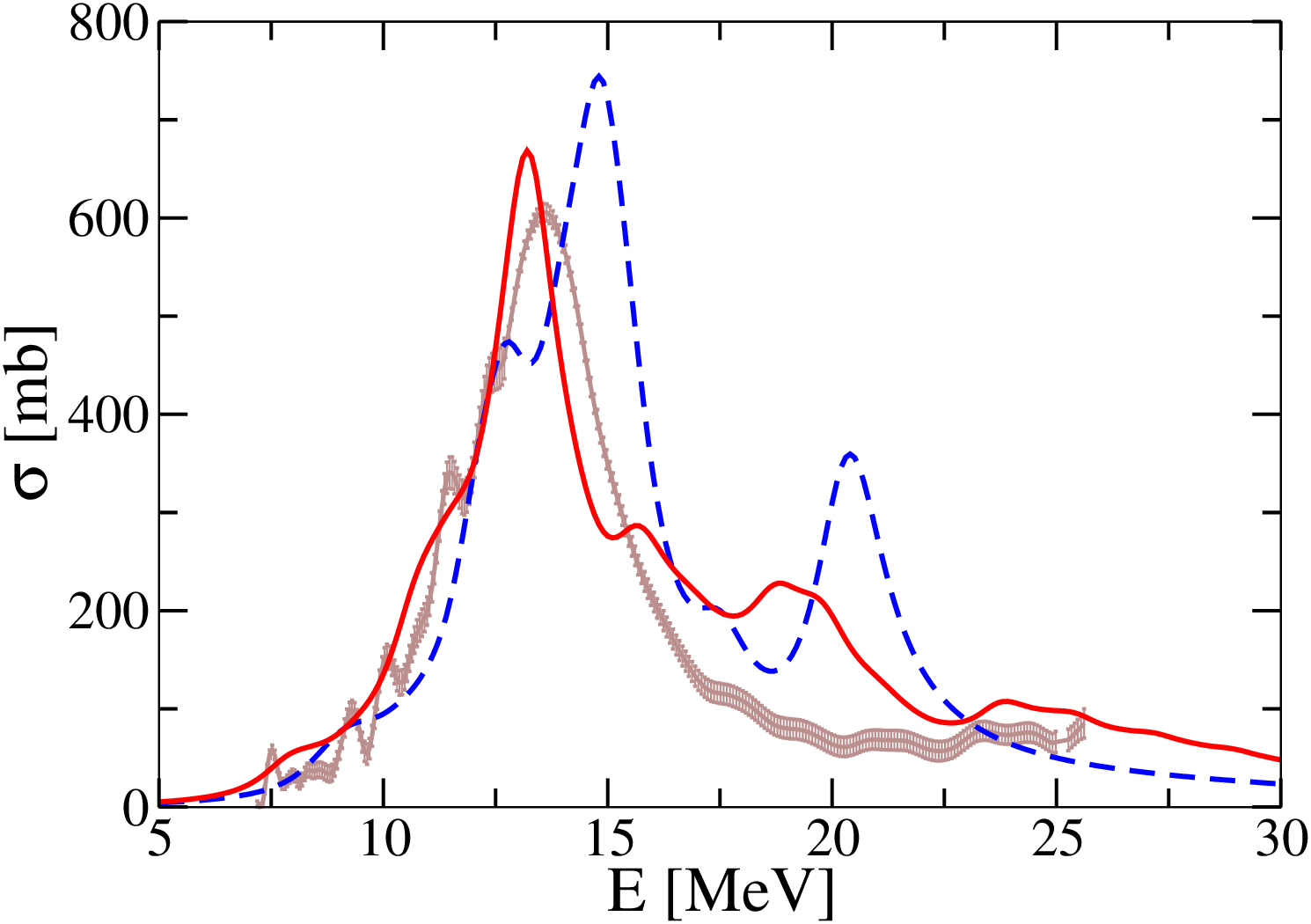}
\end{center}
\caption{\label{fig:3}(Color online) Comparison of the experimental 
photo absorption cross-section \cite{Belyaev} in $^{208}$Pb
((brown) dots with bars), with theoretical ones 
calculated in RPA (dashed(blue) line) and QTBA (solid(red) line). 
The Skyrme parametrization SV-m56-O was used. }
\end{figure}
\begin{figure}[htbp]
\vskip 4mm
\begin{center}
\includegraphics[width=8cm]{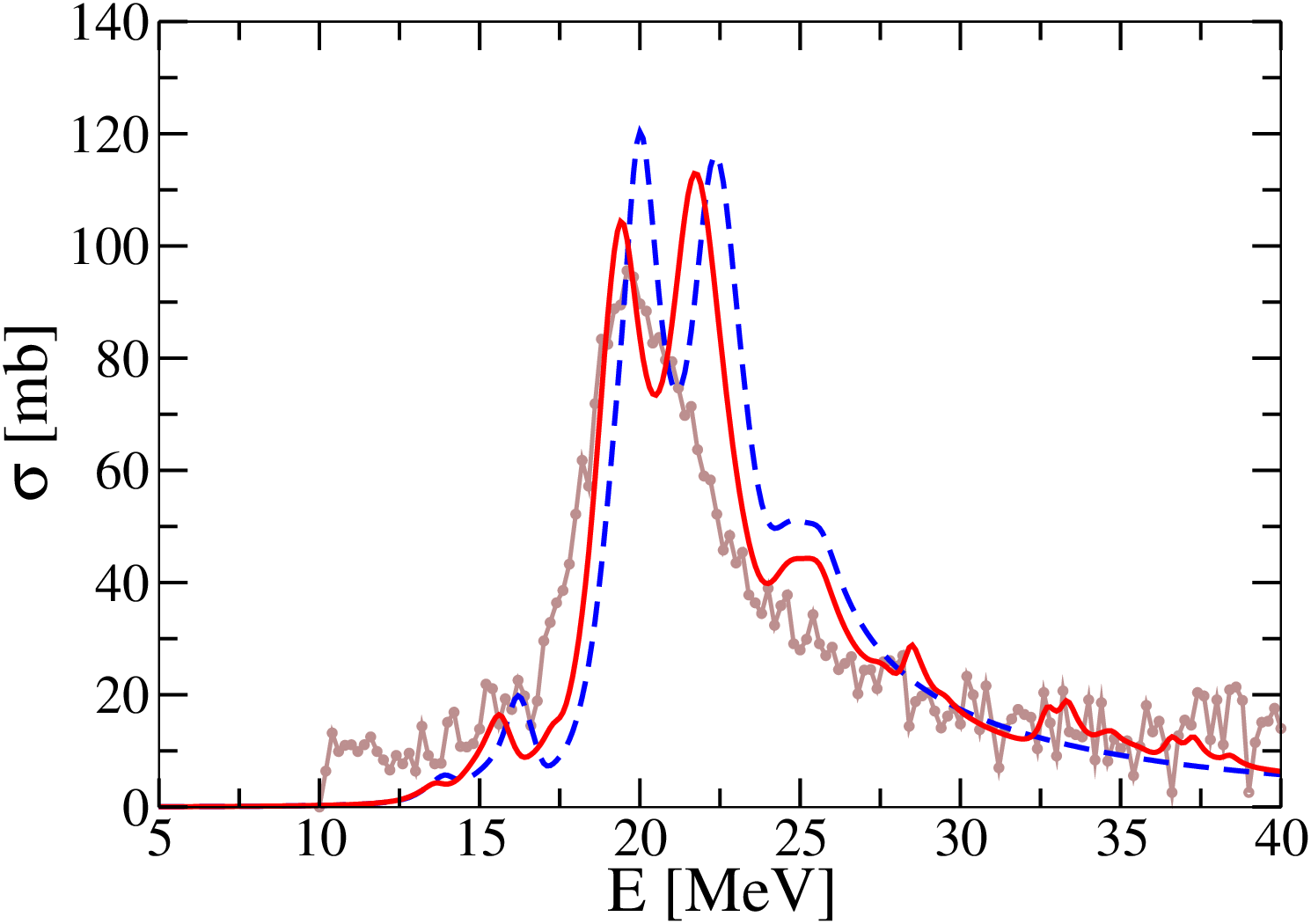}
\end{center}
\caption{\label{fig:4}(Color online) The same as in Fig. \ref{fig:3}, 
but for  $^{40}$Ca. The data are taken from Ref. \cite{Erokhova}.   } 
\end{figure}
In Fig.\ref{fig:1}, we show the sensitivity of the photo absorption 
cross sections, obtained in the framework of RPA,
on small variations of the effective mass.
We used two different values $m^*/m$=0.56 and 0.64. The higher effective mass gives  
lower GDR energies in all three nuclei. As we are interested in a Skyrme parametrization which reproduces 
the GDR in $^{16}$O we present here only results for the lower effective mass.
Note that the present calculations improve the description of the GDR 
in  $^{16}$O in comparison with other self-consistent approaches.
As only a few collective states exists in $^{16}$O, 
we do not expect strong modifications of the RPA results due to the phonons.
This is indeed the case and is demonstrated in Fig.~\ref{fig:2} 
where we compare the RPA and QTBA results. 
The Lorentzian parameters of the photo absorption cross-section \cite{Berman:1975tt} 
derived from the data of Refs.~\cite{Belyaev,Erokhova,Ishkhanov:2002}
are summarized in Table~\ref{tab:Lorentz} for $^{208}$Pb, $^{40}$Ca, 
and $^{16}$O.
The data shown here and in the subsequent figures \cite{Ishkhanov:2002,Belyaev,Erokhova}
are also available electronically\cite{nndc}.
In Fig. \ref{fig:3} we present the dipole photo absorption cross-section 
in $^{208}$Pb calculated with the Skyrme parametrization
SV-m56-O and compare them with the data \cite{Belyaev}. The result of the conventional 
RPA is compared with the QTBA where the phonons are included. 
In RPA the mean energy of the GDR  $\overline{E}$= 14.30 MeV is too high.
The rather large width $\Gamma$=4.96 MeV in the RPA is explained by a strong
peak in the cross-section at 20.4 MeV, i.e. in the high-energy tail of the GDR.
The phonons shift the GDR to lower energies, where the mean energy $\overline{E}$ = 13.37 MeV
and the width $\Gamma$=5.99 MeV are now in good agreement with the experimental data.
 We investigated the photo absorption cross section in $^{40}$Ca with the same Skyrme 
 force SV-m56-O as an example for an intermediate mass nucleus which is shown 
 in Fig.~\ref{fig:4}. The RPA result is about 1.6 MeV higher compared to the data.
The cross-section calculated within the QTBA is shifted by 0.5 MeV to lower energies 
and agrees better with experiment.
%The Lorentzian parameters of the photo absorption cross-section\cite{BerFu75} derived from the data of Refs.\cite{Belyaev,Erokhova,Ishkhanov}
%are summarized in Table \ref{tab:3n} for  $^{208}$Pb,  $^{40}$Ca, and $^{16}$O.

%\section{Summary and Outlook}
In summary, we show that the explicit inclusion of quasiparticle-phonon coupling  may solve the problem 
of mean-field theories in reproducing the  centroid energies of the giant 
dipole resonance. As the phonon contribution is small in light nuclei, 
but large in heavy mass nuclei, phonon excitations  provide a {\em mass-dependent 
 mechanism} for damping and energy shift. First  calculations  employing two new Skyrme interactions 
show a reasonable quantitative agreement with the experimental dipole excitations 
in $^{16}$O, $^{40}$Ca, and $^{208}$Pb.

%\begin{acknowledgments}
One of us (JS) thanks Stanislaw Dro\.zd\.z for many discussions and 
the Foundation for Polish Science for financial support through 
the \emph{Alexander von Humboldt Honorary Research Fellowship}.
The work was also supported by the DFG (Grant no. 436 RUS 113/994/0-1)
and RFBR (Grant no. 09-02-91352-DFG-a).

%\end{acknowledgments}
\bibliography{lyutorovich}

\begin{thebibliography}{31}%
\makeatletter
\providecommand \@ifxundefined [1]{%
 \@ifx{#1\undefined}
}%
\providecommand \@ifnum [1]{%
 \ifnum #1\expandafter \@firstoftwo
 \else \expandafter \@secondoftwo
 \fi
}%
\providecommand \@ifx [1]{%
 \ifx #1\expandafter \@firstoftwo
 \else \expandafter \@secondoftwo
 \fi
}%
\providecommand \natexlab [1]{#1}%
\providecommand \enquote  [1]{``#1''}%
\providecommand \bibnamefont  [1]{#1}%
\providecommand \bibfnamefont [1]{#1}%
\providecommand \citenamefont [1]{#1}%
\providecommand \href@noop [0]{\@secondoftwo}%
\providecommand \href [0]{\begingroup \@sanitize@url \@href}%
\providecommand \@href[1]{\@@startlink{#1}\@@href}%
\providecommand \@@href[1]{\endgroup#1\@@endlink}%
\providecommand \@sanitize@url [0]{\catcode `\\12\catcode `\$12\catcode
  `\&12\catcode `\#12\catcode `\^12\catcode `\_12\catcode `\%12\relax}%
\providecommand \@@startlink[1]{}%
\providecommand \@@endlink[0]{}%
\providecommand \url  [0]{\begingroup\@sanitize@url \@url }%
\providecommand \@url [1]{\endgroup\@href {#1}{\urlprefix }}%
\providecommand \urlprefix  [0]{URL }%
\providecommand \Eprint [0]{\href }%
\providecommand \doibase [0]{http://dx.doi.org/}%
\providecommand \selectlanguage [0]{\@gobble}%
\providecommand \bibinfo  [0]{\@secondoftwo}%
\providecommand \bibfield  [0]{\@secondoftwo}%
\providecommand \translation [1]{[#1]}%
\providecommand \BibitemOpen [0]{}%
\providecommand \bibitemStop [0]{}%
\providecommand \bibitemNoStop [0]{.\EOS\space}%
\providecommand \EOS [0]{\spacefactor3000\relax}%
\providecommand \BibitemShut  [1]{\csname bibitem#1\endcsname}%
\let\auto@bib@innerbib\@empty
%</preamble>
\bibitem [{\citenamefont {Berman}\ and\ \citenamefont
  {Fultz}(1975)}]{Berman:1975tt}%
  \BibitemOpen
  \bibfield  {author} {\bibinfo {author} {\bibfnamefont {B.}~\bibnamefont
  {Berman}}\ and\ \bibinfo {author} {\bibfnamefont {S.}~\bibnamefont {Fultz}},\
  }\href {\doibase 10.1103/RevModPhys.47.713} {\bibfield  {journal} {\bibinfo
  {journal} {Rev.Mod.Phys.}\ }\textbf {\bibinfo {volume} {47}},\ \bibinfo
  {pages} {713} (\bibinfo {year} {1975})}\BibitemShut {NoStop}%
%%CITATION = RMPHA,47,713;%%
\bibitem [{\citenamefont {Nik\v{s}i{\'{c}}}\ \emph {et~al.}(2011)\citenamefont
  {Nik\v{s}i{\'{c}}}, \citenamefont {Vretenar},\ and\ \citenamefont
  {Ring}}]{Niksic:2011sg}%
  \BibitemOpen
  \bibfield  {author} {\bibinfo {author} {\bibfnamefont {T.}~\bibnamefont
  {Nik\v{s}i{\'{c}}}}, \bibinfo {author} {\bibfnamefont {D.}~\bibnamefont
  {Vretenar}}, \ and\ \bibinfo {author} {\bibfnamefont {P.}~\bibnamefont
  {Ring}},\ }\href {\doibase 10.1016/j.ppnp.2011.01.055} {\bibfield  {journal}
  {\bibinfo  {journal} {Prog.Part.Nucl.Phys.}\ }\textbf {\bibinfo {volume}
  {66}},\ \bibinfo {pages} {519} (\bibinfo {year} {2011})},\ \Eprint
  {http://arxiv.org/abs/1102.4193} {arXiv:1102.4193 [nucl-th]} \BibitemShut
  {NoStop}%
%%CITATION = ARXIV:1102.4193;%%
\bibitem [{\citenamefont {Bender}\ \emph {et~al.}(2003)\citenamefont {Bender},
  \citenamefont {Heenen},\ and\ \citenamefont {Reinhard}}]{Bender:2003jk}%
  \BibitemOpen
  \bibfield  {author} {\bibinfo {author} {\bibfnamefont {M.}~\bibnamefont
  {Bender}}, \bibinfo {author} {\bibfnamefont {P.-H.}\ \bibnamefont {Heenen}},
  \ and\ \bibinfo {author} {\bibfnamefont {P.-G.}\ \bibnamefont {Reinhard}},\
  }\href {\doibase 10.1103/RevModPhys.75.121} {\bibfield  {journal} {\bibinfo
  {journal} {Rev.Mod.Phys.}\ }\textbf {\bibinfo {volume} {75}},\ \bibinfo
  {pages} {121} (\bibinfo {year} {2003})}\BibitemShut {NoStop}%
%%CITATION = RMPHA,75,121;%%
\bibitem [{\citenamefont {Goriely}\ \emph {et~al.}(2002)\citenamefont
  {Goriely}, \citenamefont {Samyn}, \citenamefont {Heenen}, \citenamefont
  {Pearson},\ and\ \citenamefont {Tondeur}}]{Goriely:2002xz}%
  \BibitemOpen
  \bibfield  {author} {\bibinfo {author} {\bibfnamefont {S.}~\bibnamefont
  {Goriely}}, \bibinfo {author} {\bibfnamefont {M.}~\bibnamefont {Samyn}},
  \bibinfo {author} {\bibfnamefont {P.-H.}\ \bibnamefont {Heenen}}, \bibinfo
  {author} {\bibfnamefont {J.}~\bibnamefont {Pearson}}, \ and\ \bibinfo
  {author} {\bibfnamefont {F.}~\bibnamefont {Tondeur}},\ }\href {\doibase
  10.1103/PhysRevC.66.024326} {\bibfield  {journal} {\bibinfo  {journal}
  {Phys.Rev.}\ }\textbf {\bibinfo {volume} {C66}},\ \bibinfo {pages} {024326}
  (\bibinfo {year} {2002})}\BibitemShut {NoStop}%
%%CITATION = PHRVA,C66,024326;%%
\bibitem [{\citenamefont {Kortelainen}\ \emph {et~al.}(2010)\citenamefont
  {Kortelainen}, \citenamefont {Lesinski}, \citenamefont {Mor{\'{e}}},
  \citenamefont {Nazarewicz}, \citenamefont {Sarich} \emph
  {et~al.}}]{Kortelainen:2010hv}%
  \BibitemOpen
  \bibfield  {author} {\bibinfo {author} {\bibfnamefont {M.}~\bibnamefont
  {Kortelainen}}, \bibinfo {author} {\bibfnamefont {T.}~\bibnamefont
  {Lesinski}}, \bibinfo {author} {\bibfnamefont {J.}~\bibnamefont
  {Mor{\'{e}}}}, \bibinfo {author} {\bibfnamefont {W.}~\bibnamefont
  {Nazarewicz}}, \bibinfo {author} {\bibfnamefont {J.}~\bibnamefont {Sarich}},
  \emph {et~al.},\ }\href {\doibase 10.1103/PhysRevC.82.024313} {\bibfield
  {journal} {\bibinfo  {journal} {Phys.Rev.}\ }\textbf {\bibinfo {volume}
  {C82}},\ \bibinfo {pages} {024313} (\bibinfo {year} {2010})},\ \Eprint
  {http://arxiv.org/abs/1005.5145} {arXiv:1005.5145 [nucl-th]} \BibitemShut
  {NoStop}%
%%CITATION = ARXIV:1005.5145;%%
\bibitem [{\citenamefont {Kl\"upfel}\ \emph {et~al.}(2009)\citenamefont
  {Kl\"upfel}, \citenamefont {Reinhard}, \citenamefont {B\"urvenich},\ and\
  \citenamefont {Maruhn}}]{Klupfel:2008af}%
  \BibitemOpen
  \bibfield  {author} {\bibinfo {author} {\bibfnamefont {P.}~\bibnamefont
  {Kl\"upfel}}, \bibinfo {author} {\bibfnamefont {P.-G.}\ \bibnamefont
  {Reinhard}}, \bibinfo {author} {\bibfnamefont {T.}~\bibnamefont
  {B\"urvenich}}, \ and\ \bibinfo {author} {\bibfnamefont {J.}~\bibnamefont
  {Maruhn}},\ }\href {\doibase 10.1103/PhysRevC.79.034310} {\bibfield
  {journal} {\bibinfo  {journal} {Phys.Rev.}\ }\textbf {\bibinfo {volume}
  {C79}},\ \bibinfo {pages} {034310} (\bibinfo {year} {2009})},\ \Eprint
  {http://arxiv.org/abs/0804.3385} {arXiv:0804.3385 [nucl-th]} \BibitemShut
  {NoStop}%
%%CITATION = ARXIV:0804.3385;%%
\bibitem [{\citenamefont {Erler}\ \emph {et~al.}(2011)\citenamefont {Erler},
  \citenamefont {Kl\"upfel},\ and\ \citenamefont {Reinhard}}]{Erler:2011}%
  \BibitemOpen
  \bibfield  {author} {\bibinfo {author} {\bibfnamefont {J.}~\bibnamefont
  {Erler}}, \bibinfo {author} {\bibfnamefont {P.}~\bibnamefont {Kl\"upfel}}, \
  and\ \bibinfo {author} {\bibfnamefont {P.-G.}\ \bibnamefont {Reinhard}},\
  }\href {http://stacks.iop.org/0954-3899/38/i=3/a=033101} {\bibfield
  {journal} {\bibinfo  {journal} {Journal of Physics G: Nuclear and Particle
  Physics}\ }\textbf {\bibinfo {volume} {38}},\ \bibinfo {pages} {033101}
  (\bibinfo {year} {2011})}\BibitemShut {NoStop}%
\bibitem [{\citenamefont {Erler}\ \emph {et~al.}(2010)\citenamefont {Erler},
  \citenamefont {Kl\"upfel},\ and\ \citenamefont {Reinhard}}]{Erler:2010fe}%
  \BibitemOpen
  \bibfield  {author} {\bibinfo {author} {\bibfnamefont {J.}~\bibnamefont
  {Erler}}, \bibinfo {author} {\bibfnamefont {P.}~\bibnamefont {Kl\"upfel}}, \
  and\ \bibinfo {author} {\bibfnamefont {P.-G.}\ \bibnamefont {Reinhard}},\
  }\href {\doibase 10.1088/0954-3899/37/6/064001} {\bibfield  {journal}
  {\bibinfo  {journal} {J.Phys.G}\ }\textbf {\bibinfo {volume} {37}},\ \bibinfo
  {pages} {064001} (\bibinfo {year} {2010})},\ \Eprint
  {http://arxiv.org/abs/1002.0027} {arXiv:1002.0027 [nucl-th]} \BibitemShut
  {NoStop}%
%%CITATION = ARXIV:1002.0027;%%
\bibitem [{\citenamefont {Dutra}\ \emph {et~al.}(2012)\citenamefont {Dutra},
  \citenamefont {Lourenco}, \citenamefont {S{\'{a}}~Martins}, \citenamefont
  {Delfino}, \citenamefont {Stone},\ and\ \citenamefont
  {Stevenson}}]{Dutra:2012mb}%
  \BibitemOpen
  \bibfield  {author} {\bibinfo {author} {\bibfnamefont {M.}~\bibnamefont
  {Dutra}}, \bibinfo {author} {\bibfnamefont {O.}~\bibnamefont {Lourenco}},
  \bibinfo {author} {\bibfnamefont {J.}~\bibnamefont {S{\'{a}}~Martins}},
  \bibinfo {author} {\bibfnamefont {A.}~\bibnamefont {Delfino}}, \bibinfo
  {author} {\bibfnamefont {J.}~\bibnamefont {Stone}}, \ and\ \bibinfo {author}
  {\bibfnamefont {P.}~\bibnamefont {Stevenson}},\ }\href {\doibase
  10.1103/PhysRevC.85.035201} {\bibfield  {journal} {\bibinfo  {journal}
  {Phys.Rev.}\ }\textbf {\bibinfo {volume} {C85}},\ \bibinfo {pages} {035201}
  (\bibinfo {year} {2012})},\ \Eprint {http://arxiv.org/abs/1202.3902}
  {arXiv:1202.3902 [nucl-th]} \BibitemShut {NoStop}%
%%CITATION = ARXIV:1202.3902;%%
\bibitem [{\citenamefont {Abrahamyan}\ \emph {et~al.}(2012)\citenamefont
  {Abrahamyan}, \citenamefont {Ahmed} \emph {et~al.}}]{Abrahamyan:2012}%
  \BibitemOpen
  \bibfield  {author} {\bibinfo {author} {\bibfnamefont {S.}~\bibnamefont
  {Abrahamyan}}, \bibinfo {author} {\bibfnamefont {Z.}~\bibnamefont {Ahmed}},
  \emph {et~al.} (\bibinfo {collaboration} {PREX Collaboration}),\ }\href
  {\doibase 10.1103/PhysRevLett.108.112502} {\bibfield  {journal} {\bibinfo
  {journal} {Phys. Rev. Lett.}\ }\textbf {\bibinfo {volume} {108}},\ \bibinfo
  {pages} {112502} (\bibinfo {year} {2012})}\BibitemShut {NoStop}%
\bibitem [{\citenamefont {Tamii}\ \emph {et~al.}(2011)\citenamefont {Tamii},
  \citenamefont {Poltoratska}, \citenamefont {von Neumann-Cosel}, \citenamefont
  {Fujita}, \citenamefont {Adachi} \emph {et~al.}}]{Tamii:2011pv}%
  \BibitemOpen
  \bibfield  {author} {\bibinfo {author} {\bibfnamefont {A.}~\bibnamefont
  {Tamii}}, \bibinfo {author} {\bibfnamefont {I.}~\bibnamefont {Poltoratska}},
  \bibinfo {author} {\bibfnamefont {P.}~\bibnamefont {von Neumann-Cosel}},
  \bibinfo {author} {\bibfnamefont {Y.}~\bibnamefont {Fujita}}, \bibinfo
  {author} {\bibfnamefont {T.}~\bibnamefont {Adachi}},  \emph {et~al.},\ }\href
  {\doibase 10.1103/PhysRevLett.107.062502} {\bibfield  {journal} {\bibinfo
  {journal} {Phys.Rev.Lett.}\ }\textbf {\bibinfo {volume} {107}},\ \bibinfo
  {pages} {062502} (\bibinfo {year} {2011})},\ \Eprint
  {http://arxiv.org/abs/1104.5431} {arXiv:1104.5431 [nucl-ex]} \BibitemShut
  {NoStop}%
%%CITATION = ARXIV:1104.5431;%%
\bibitem [{\citenamefont {Savran}\ \emph {et~al.}(2011)\citenamefont {Savran},
  \citenamefont {Elvers}, \citenamefont {Endres}, \citenamefont {Fritzsche},
  \citenamefont {L\"oher} \emph {et~al.}}]{Savran:2011zza}%
  \BibitemOpen
  \bibfield  {author} {\bibinfo {author} {\bibfnamefont {D.}~\bibnamefont
  {Savran}}, \bibinfo {author} {\bibfnamefont {M.}~\bibnamefont {Elvers}},
  \bibinfo {author} {\bibfnamefont {J.}~\bibnamefont {Endres}}, \bibinfo
  {author} {\bibfnamefont {M.}~\bibnamefont {Fritzsche}}, \bibinfo {author}
  {\bibfnamefont {B.}~\bibnamefont {L\"oher}},  \emph {et~al.},\ }\href
  {\doibase 10.1103/PhysRevC.84.024326} {\bibfield  {journal} {\bibinfo
  {journal} {Phys.Rev.}\ }\textbf {\bibinfo {volume} {C84}},\ \bibinfo {pages}
  {024326} (\bibinfo {year} {2011})}\BibitemShut {NoStop}%
%%CITATION = PHRVA,C84,024326;%%
\bibitem [{\citenamefont {Horowitz}\ and\ \citenamefont
  {Piekarewicz}(2001)}]{Horowitz:2000xj}%
  \BibitemOpen
  \bibfield  {author} {\bibinfo {author} {\bibfnamefont {C.}~\bibnamefont
  {Horowitz}}\ and\ \bibinfo {author} {\bibfnamefont {J.}~\bibnamefont
  {Piekarewicz}},\ }\href {\doibase 10.1103/PhysRevLett.86.5647} {\bibfield
  {journal} {\bibinfo  {journal} {Phys.Rev.Lett.}\ }\textbf {\bibinfo {volume}
  {86}},\ \bibinfo {pages} {5647} (\bibinfo {year} {2001})},\ \Eprint
  {http://arxiv.org/abs/astro-ph/0010227} {arXiv:astro-ph/0010227 [astro-ph]}
  \BibitemShut {NoStop}%
%%CITATION = ASTRO-PH/0010227;%%
\bibitem [{\citenamefont {Bertsch}\ \emph {et~al.}(1983)\citenamefont
  {Bertsch}, \citenamefont {Bortignon},\ and\ \citenamefont
  {Broglia}}]{Bertsch:1983zz}%
  \BibitemOpen
  \bibfield  {author} {\bibinfo {author} {\bibfnamefont {G.}~\bibnamefont
  {Bertsch}}, \bibinfo {author} {\bibfnamefont {P.}~\bibnamefont {Bortignon}},
  \ and\ \bibinfo {author} {\bibfnamefont {R.}~\bibnamefont {Broglia}},\ }\href
  {\doibase 10.1103/RevModPhys.55.287} {\bibfield  {journal} {\bibinfo
  {journal} {Rev.Mod.Phys.}\ }\textbf {\bibinfo {volume} {55}},\ \bibinfo
  {pages} {287} (\bibinfo {year} {1983})}\BibitemShut {NoStop}%
%%CITATION = RMPHA,55,287;%%
\bibitem [{\citenamefont {Tselyaev}(1989)}]{Tselyaev:1989}%
  \BibitemOpen
  \bibfield  {author} {\bibinfo {author} {\bibfnamefont {V.}~\bibnamefont
  {Tselyaev}},\ }\href@noop {} {\bibfield  {journal} {\bibinfo  {journal}
  {Yad.Fiz.; Soviet Journal of Nuclear Physics (English translation)}\ }\textbf
  {\bibinfo {volume} {50}},\ \bibinfo {pages} {1252} (\bibinfo {year}
  {1989})}\BibitemShut {NoStop}%
\bibitem [{\citenamefont {Tselyaev}(2007)}]{Tselyaev:2005de}%
  \BibitemOpen
  \bibfield  {author} {\bibinfo {author} {\bibfnamefont {V.}~\bibnamefont
  {Tselyaev}},\ }\href {\doibase 10.1103/PhysRevC.75.024306} {\bibfield
  {journal} {\bibinfo  {journal} {Phys.Rev.}\ }\textbf {\bibinfo {volume}
  {C75}},\ \bibinfo {pages} {024306} (\bibinfo {year} {2007})},\ \Eprint
  {http://arxiv.org/abs/nucl-th/0505031} {arXiv:nucl-th/0505031 [nucl-th]}
  \BibitemShut {NoStop}%
%%CITATION = NUCL-TH/0505031;%%
\bibitem [{\citenamefont {Kamerdzhiev}\ \emph {et~al.}(1993)\citenamefont
  {Kamerdzhiev}, \citenamefont {Speth}, \citenamefont {Tertychny},\ and\
  \citenamefont {Tselyaev}}]{Kamerdzhiev199390}%
  \BibitemOpen
  \bibfield  {author} {\bibinfo {author} {\bibfnamefont {S.}~\bibnamefont
  {Kamerdzhiev}}, \bibinfo {author} {\bibfnamefont {J.}~\bibnamefont {Speth}},
  \bibinfo {author} {\bibfnamefont {G.}~\bibnamefont {Tertychny}}, \ and\
  \bibinfo {author} {\bibfnamefont {V.}~\bibnamefont {Tselyaev}},\ }\href
  {\doibase 10.1016/0375-9474(93)90315-O} {\bibfield  {journal} {\bibinfo
  {journal} {Nuclear Physics A}\ }\textbf {\bibinfo {volume} {555}},\ \bibinfo
  {pages} {90 } (\bibinfo {year} {1993})}\BibitemShut {NoStop}%
\bibitem [{\citenamefont {Avdeenkov}\ \emph {et~al.}(2007)\citenamefont
  {Avdeenkov}, \citenamefont {Gr\"ummer}, \citenamefont {Kamerdzhiev},
  \citenamefont {Krewald}, \citenamefont {Lyutorovich},\ and\ \citenamefont
  {Speth}}]{Avdeenkov2007196}%
  \BibitemOpen
  \bibfield  {author} {\bibinfo {author} {\bibfnamefont {A.}~\bibnamefont
  {Avdeenkov}}, \bibinfo {author} {\bibfnamefont {F.}~\bibnamefont
  {Gr\"ummer}}, \bibinfo {author} {\bibfnamefont {S.}~\bibnamefont
  {Kamerdzhiev}}, \bibinfo {author} {\bibfnamefont {S.}~\bibnamefont
  {Krewald}}, \bibinfo {author} {\bibfnamefont {N.}~\bibnamefont
  {Lyutorovich}}, \ and\ \bibinfo {author} {\bibfnamefont {J.}~\bibnamefont
  {Speth}},\ }\href {\doibase 10.1016/j.physletb.2007.07.069} {\bibfield
  {journal} {\bibinfo  {journal} {Physics Letters B}\ }\textbf {\bibinfo
  {volume} {653}},\ \bibinfo {pages} {196 } (\bibinfo {year}
  {2007})}\BibitemShut {NoStop}%
\bibitem [{\citenamefont {Lyutorovich}\ \emph {et~al.}(2008)\citenamefont
  {Lyutorovich}, \citenamefont {Speth}, \citenamefont {Avdeenkov},
  \citenamefont {Gr\"ummer}, \citenamefont {Kamerdzhiev} \emph
  {et~al.}}]{Lyutorovich:2008nv}%
  \BibitemOpen
  \bibfield  {author} {\bibinfo {author} {\bibfnamefont {N.}~\bibnamefont
  {Lyutorovich}}, \bibinfo {author} {\bibfnamefont {J.}~\bibnamefont {Speth}},
  \bibinfo {author} {\bibfnamefont {A.}~\bibnamefont {Avdeenkov}}, \bibinfo
  {author} {\bibfnamefont {F.}~\bibnamefont {Gr\"ummer}}, \bibinfo {author}
  {\bibfnamefont {S.}~\bibnamefont {Kamerdzhiev}},  \emph {et~al.},\ }\href
  {\doibase 10.1140/epja/i2008-10638-x} {\bibfield  {journal} {\bibinfo
  {journal} {Eur.Phys.J.}\ }\textbf {\bibinfo {volume} {A37}},\ \bibinfo
  {pages} {381} (\bibinfo {year} {2008})},\ \Eprint
  {http://arxiv.org/abs/0806.2813} {arXiv:0806.2813 [nucl-th]} \BibitemShut
  {NoStop}%
%%CITATION = ARXIV:0806.2813;%%
\bibitem [{\citenamefont {Tselyaev}\ \emph {et~al.}(2007)\citenamefont
  {Tselyaev}, \citenamefont {Speth}, \citenamefont {Gr\"ummer}, \citenamefont
  {Krewald}, \citenamefont {Avdeenkov}, \citenamefont {Litvinova},\ and\
  \citenamefont {Tertychny}}]{Tselyaev_Speth:2007}%
  \BibitemOpen
  \bibfield  {author} {\bibinfo {author} {\bibfnamefont {V.}~\bibnamefont
  {Tselyaev}}, \bibinfo {author} {\bibfnamefont {J.}~\bibnamefont {Speth}},
  \bibinfo {author} {\bibfnamefont {F.}~\bibnamefont {Gr\"ummer}}, \bibinfo
  {author} {\bibfnamefont {S.}~\bibnamefont {Krewald}}, \bibinfo {author}
  {\bibfnamefont {A.}~\bibnamefont {Avdeenkov}}, \bibinfo {author}
  {\bibfnamefont {E.}~\bibnamefont {Litvinova}}, \ and\ \bibinfo {author}
  {\bibfnamefont {G.}~\bibnamefont {Tertychny}},\ }\href {\doibase
  10.1103/PhysRevC.75.014315} {\bibfield  {journal} {\bibinfo  {journal} {Phys.
  Rev. C}\ }\textbf {\bibinfo {volume} {75}},\ \bibinfo {pages} {014315}
  (\bibinfo {year} {2007})}\BibitemShut {NoStop}%
\bibitem [{\citenamefont {Kamerdzhiev}\ \emph {et~al.}(2004)\citenamefont
  {Kamerdzhiev}, \citenamefont {Speth},\ and\ \citenamefont
  {Tertychny}}]{Kamerdzhiev:2003rd}%
  \BibitemOpen
  \bibfield  {author} {\bibinfo {author} {\bibfnamefont {S.}~\bibnamefont
  {Kamerdzhiev}}, \bibinfo {author} {\bibfnamefont {J.}~\bibnamefont {Speth}},
  \ and\ \bibinfo {author} {\bibfnamefont {G.}~\bibnamefont {Tertychny}},\
  }\href {\doibase 10.1016/j.physrep.2003.11.001} {\bibfield  {journal}
  {\bibinfo  {journal} {Phys.Rept.}\ }\textbf {\bibinfo {volume} {393}},\
  \bibinfo {pages} {1} (\bibinfo {year} {2004})},\ \Eprint
  {http://arxiv.org/abs/nucl-th/0311058} {arXiv:nucl-th/0311058 [nucl-th]}
  \BibitemShut {NoStop}%
%%CITATION = NUCL-TH/0311058;%%
\bibitem [{\citenamefont {Tselyaev}\ \emph {et~al.}(2009)\citenamefont
  {Tselyaev}, \citenamefont {Speth}, \citenamefont {Gr\"ummer}, \citenamefont
  {Krewald}, \citenamefont {Avdeenkov}, \citenamefont {Litvinova},\ and\
  \citenamefont {Lyutorovich}}]{Tselyaev2009}%
  \BibitemOpen
  \bibfield  {author} {\bibinfo {author} {\bibfnamefont {V.}~\bibnamefont
  {Tselyaev}}, \bibinfo {author} {\bibfnamefont {J.}~\bibnamefont {Speth}},
  \bibinfo {author} {\bibfnamefont {F.}~\bibnamefont {Gr\"ummer}}, \bibinfo
  {author} {\bibfnamefont {S.}~\bibnamefont {Krewald}}, \bibinfo {author}
  {\bibfnamefont {A.}~\bibnamefont {Avdeenkov}}, \bibinfo {author}
  {\bibfnamefont {E.}~\bibnamefont {Litvinova}}, \ and\ \bibinfo {author}
  {\bibfnamefont {A.}~\bibnamefont {Lyutorovich}},\ }\href {\doibase
  10.1103/PhysRevC.79.034309} {\bibfield  {journal} {\bibinfo  {journal} {Phys.
  Rev. C}\ }\textbf {\bibinfo {volume} {79}},\ \bibinfo {pages} {034309}
  (\bibinfo {year} {2009})}\BibitemShut {NoStop}%
\bibitem [{\citenamefont {Brown}\ \emph {et~al.}(1963)\citenamefont {Brown},
  \citenamefont {Gunn},\ and\ \citenamefont {Gould}}]{Brown1963598}%
  \BibitemOpen
  \bibfield  {author} {\bibinfo {author} {\bibfnamefont {G.}~\bibnamefont
  {Brown}}, \bibinfo {author} {\bibfnamefont {J.}~\bibnamefont {Gunn}}, \ and\
  \bibinfo {author} {\bibfnamefont {P.}~\bibnamefont {Gould}},\ }\href
  {\doibase 10.1016/0029-5582(63)90631-X} {\bibfield  {journal} {\bibinfo
  {journal} {Nuclear Physics}\ }\textbf {\bibinfo {volume} {46}},\ \bibinfo
  {pages} {598 } (\bibinfo {year} {1963})}\BibitemShut {NoStop}%
\bibitem [{\citenamefont {Bertsch}\ and\ \citenamefont
  {Kuo}(1968)}]{Bertsch1968204}%
  \BibitemOpen
  \bibfield  {author} {\bibinfo {author} {\bibfnamefont {G.}~\bibnamefont
  {Bertsch}}\ and\ \bibinfo {author} {\bibfnamefont {T.}~\bibnamefont {Kuo}},\
  }\href {\doibase 10.1016/0375-9474(68)90230-3} {\bibfield  {journal}
  {\bibinfo  {journal} {Nuclear Physics A}\ }\textbf {\bibinfo {volume}
  {112}},\ \bibinfo {pages} {204 } (\bibinfo {year} {1968})}\BibitemShut
  {NoStop}%
\bibitem [{\citenamefont {Ring}\ and\ \citenamefont
  {Werner}(1973)}]{Ring1973198}%
  \BibitemOpen
  \bibfield  {author} {\bibinfo {author} {\bibfnamefont {P.}~\bibnamefont
  {Ring}}\ and\ \bibinfo {author} {\bibfnamefont {E.}~\bibnamefont {Werner}},\
  }\href {\doibase 10.1016/0375-9474(73)90773-2} {\bibfield  {journal}
  {\bibinfo  {journal} {Nuclear Physics A}\ }\textbf {\bibinfo {volume}
  {211}},\ \bibinfo {pages} {198 } (\bibinfo {year} {1973})}\BibitemShut
  {NoStop}%
\bibitem [{\citenamefont {Jeukenne}\ \emph {et~al.}(1976)\citenamefont
  {Jeukenne}, \citenamefont {Lejeune},\ and\ \citenamefont
  {Mahaux}}]{Jeukenne:1976uy}%
  \BibitemOpen
  \bibfield  {author} {\bibinfo {author} {\bibfnamefont {J.}~\bibnamefont
  {Jeukenne}}, \bibinfo {author} {\bibfnamefont {A.}~\bibnamefont {Lejeune}}, \
  and\ \bibinfo {author} {\bibfnamefont {C.}~\bibnamefont {Mahaux}},\ }\href
  {\doibase 10.1016/0370-1573(76)90017-X} {\bibfield  {journal} {\bibinfo
  {journal} {Phys.Rept.}\ }\textbf {\bibinfo {volume} {25}},\ \bibinfo {pages}
  {83} (\bibinfo {year} {1976})}\BibitemShut {NoStop}%
%%CITATION = PRPLC,25,83;%%
\bibitem [{\citenamefont {Lejeune}\ and\ \citenamefont
  {Mahaux}(1981)}]{Leieune:1979}%
  \BibitemOpen
  \bibfield  {author} {\bibinfo {author} {\bibfnamefont {A.}~\bibnamefont
  {Lejeune}}\ and\ \bibinfo {author} {\bibfnamefont {C.}~\bibnamefont
  {Mahaux}},\ }\enquote {\bibinfo {title} {{Proceedings Enrico Fermi School
  LXXVII,1979}},}\ \ (\bibinfo  {publisher} {North-Holland},\ \bibinfo {year}
  {1981})\ p.\ \bibinfo {pages} {418}\BibitemShut {NoStop}%
\bibitem [{\citenamefont {Ishkhanov}\ \emph {et~al.}(2002)\citenamefont
  {Ishkhanov}, \citenamefont {Kapitonov}, \citenamefont {Lileeva},
  \citenamefont {Shirokov}, \citenamefont {Erokhova}, \citenamefont {Elkin},\
  and\ \citenamefont {Izotova}}]{Ishkhanov:2002}%
  \BibitemOpen
  \bibfield  {author} {\bibinfo {author} {\bibfnamefont {B.}~\bibnamefont
  {Ishkhanov}}, \bibinfo {author} {\bibfnamefont {I.}~\bibnamefont
  {Kapitonov}}, \bibinfo {author} {\bibfnamefont {E.}~\bibnamefont {Lileeva}},
  \bibinfo {author} {\bibfnamefont {E.}~\bibnamefont {Shirokov}}, \bibinfo
  {author} {\bibfnamefont {V.}~\bibnamefont {Erokhova}}, \bibinfo {author}
  {\bibfnamefont {M.}~\bibnamefont {Elkin}}, \ and\ \bibinfo {author}
  {\bibfnamefont {A.}~\bibnamefont {Izotova}},\ }\href@noop {} {}\bibinfo
  {howpublished} {preprint MSU-INP-2002-27/711} (\bibinfo {year}
  {2002})\BibitemShut {NoStop}%
\bibitem [{\citenamefont {Belyaev}\ \emph {et~al.}(1995)\citenamefont
  {Belyaev}, \citenamefont {Vasiliev}, \citenamefont {Voronov}, \citenamefont
  {Nechkin}, \citenamefont {Ponomarev},\ and\ \citenamefont
  {Semenov}}]{Belyaev}%
  \BibitemOpen
  \bibfield  {author} {\bibinfo {author} {\bibfnamefont {S.}~\bibnamefont
  {Belyaev}}, \bibinfo {author} {\bibfnamefont {O.}~\bibnamefont {Vasiliev}},
  \bibinfo {author} {\bibfnamefont {V.}~\bibnamefont {Voronov}}, \bibinfo
  {author} {\bibfnamefont {A.}~\bibnamefont {Nechkin}}, \bibinfo {author}
  {\bibfnamefont {V.}~\bibnamefont {Ponomarev}}, \ and\ \bibinfo {author}
  {\bibfnamefont {V.}~\bibnamefont {Semenov}},\ }\href@noop {} {\bibfield
  {journal} {\bibinfo  {journal} {Phys.Atom.Nucl.}\ }\textbf {\bibinfo {volume}
  {58}},\ \bibinfo {pages} {1883} (\bibinfo {year} {1995})}\BibitemShut
  {NoStop}%
\bibitem [{\citenamefont {Erokhova}\ \emph {et~al.}(2003)\citenamefont
  {Erokhova}, \citenamefont {Elkin}, \citenamefont {Izotova}, \citenamefont
  {Ishkhanov}, \citenamefont {Kapitonov}, \citenamefont {Lileeva},\ and\
  \citenamefont {Shirokov}}]{Erokhova}%
  \BibitemOpen
  \bibfield  {author} {\bibinfo {author} {\bibfnamefont {V.}~\bibnamefont
  {Erokhova}}, \bibinfo {author} {\bibfnamefont {M.}~\bibnamefont {Elkin}},
  \bibinfo {author} {\bibfnamefont {A.}~\bibnamefont {Izotova}}, \bibinfo
  {author} {\bibfnamefont {B.}~\bibnamefont {Ishkhanov}}, \bibinfo {author}
  {\bibfnamefont {L.}~\bibnamefont {Kapitonov}}, \bibinfo {author}
  {\bibfnamefont {E.}~\bibnamefont {Lileeva}}, \ and\ \bibinfo {author}
  {\bibfnamefont {E.}~\bibnamefont {Shirokov}},\ }\href@noop {} {\bibfield
  {journal} {\bibinfo  {journal} {Izv.Ross.Akad.Nauk.Ser.Fiz..}\ }\textbf
  {\bibinfo {volume} {67}},\ \bibinfo {pages} {1479} (\bibinfo {year}
  {2003})}\BibitemShut {NoStop}%
\bibitem [{nnd()}]{nndc}%
  \BibitemOpen
  \href@noop {} {}\bibinfo {howpublished}
  {http://www.nndc.bnl.gov/exfor/exfor00.htm}\BibitemShut {NoStop}%
\end{thebibliography}%

\end{document}